\documentclass[conference, 10pt]{IEEEtran}

\IEEEoverridecommandlockouts
\usepackage{amsmath,amssymb,amsfonts}
\usepackage{graphicx}
\usepackage{textcomp}
\usepackage{xcolor}
\usepackage{algorithm}
\usepackage{algorithmicx}
\usepackage{algpseudocode}
\usepackage{subcaption}
\usepackage{booktabs}
\usepackage[final]{microtype}
\usepackage{setspace}
\usepackage{enumitem}
\usepackage{tikz}
\usepackage{times}
\usepackage{newtxtext,newtxmath}
\usepackage{comment}
\usepackage{hyperref}
\usepackage{cite}
\usepackage{cleveref}
\usepackage{wrapfig}

\newcommand{\niparagraph}[1]{\vspace{1pt}\noindent\textbf{#1}}

\setlist[itemize]{noitemsep, topsep=0pt, leftmargin=*}
\author{
    \IEEEauthorblockN{
        Sumukh Pinge\IEEEauthorrefmark{1},
        Ashkan Moradifirouzabadi\IEEEauthorrefmark{1},
        Keming Fan\IEEEauthorrefmark{1},
        Prasanna Venkatesan Ravindran\IEEEauthorrefmark{3},
        Tanvir H. Pantha\IEEEauthorrefmark{2},
    }
    \IEEEauthorblockN{
        Po\hbox{-}Kai Hsu\IEEEauthorrefmark{3},
        Zheyu Li\IEEEauthorrefmark{1},
        Weihong Xu\IEEEauthorrefmark{1},
        Zihan Xia\IEEEauthorrefmark{1},
        Flavio Ponzina\IEEEauthorrefmark{1},
        Winston Chern\IEEEauthorrefmark{3},
        Taeyoung Song\IEEEauthorrefmark{3},
    }
    \IEEEauthorblockN{
        Priyankka Ravikumar\IEEEauthorrefmark{3},
        Mengkun Tian\IEEEauthorrefmark{3},
        Lance Fernandes\IEEEauthorrefmark{3},
        Huy Tran\IEEEauthorrefmark{3},
        Hari Jayasankar\IEEEauthorrefmark{3},
        Hang Chen\IEEEauthorrefmark{3},
    }
    \IEEEauthorblockN{
        Chinsung Park\IEEEauthorrefmark{3},
        Amrit Garlapati\IEEEauthorrefmark{3},
        Kijoon Kim\IEEEauthorrefmark{4},
        Jongho Woo\IEEEauthorrefmark{4},
        Suhwan Lim\IEEEauthorrefmark{4},
        Kwangsoo Kim\IEEEauthorrefmark{4},
        Wanki Kim\IEEEauthorrefmark{4},
    }
    \IEEEauthorblockN{
        Daewon Ha\IEEEauthorrefmark{4},
        Duygu Kuzum\IEEEauthorrefmark{1},
        Shimeng Yu\IEEEauthorrefmark{3},        
        Sourav Dutta\IEEEauthorrefmark{2},
        Asif Khan\IEEEauthorrefmark{3},
        Tajana Rosing\IEEEauthorrefmark{1},
        Mingu Kang\IEEEauthorrefmark{1}
    }

    \vspace{0.1cm}
    \IEEEauthorblockA{\IEEEauthorrefmark{1}University of California San Diego, La Jolla, CA 92093, USA}
    \IEEEauthorblockA{\IEEEauthorrefmark{2}The University of Texas at Dallas, Richardson, TX 75080, USA}
    \IEEEauthorblockA{\IEEEauthorrefmark{3}Georgia Institute of Technology, Atlanta, GA 30332, USA}
    \IEEEauthorblockA{\IEEEauthorrefmark{4}Advanced Device Research Laboratory, Semiconductor Research Center, Samsung Electronics Co., Ltd., South Korea}
    \IEEEauthorblockA{\href{mailto:{spinge, tajana, m7kang}@ucsd.edu}{\{spinge, tajana, mingu\}@ucsd.edu}}
}

\def\BibTeX{{\rm B\kern-.05em{\sc i\kern-.025em b}\kern-.08em
    T\kern-.1667em\lower.7ex\hbox{E}\kern-.125emX}}
\begin{document}

\title{FeNOMS: Enhancing Open Modification Spectral Library Search with In-Storage Processing on Ferroelectric NAND (FeNAND) Flash \vspace*{-0.5cm}}

\maketitle
\begin{abstract}
The rapid expansion of mass spectrometry (MS) data, now exceeding hundreds of terabytes, poses significant challenges for efficient, large-scale library search — a critical component for drug discovery. Traditional processors struggle to handle this data volume efficiently, making in-storage computing (ISP) a promising alternative. This work introduces an ISP architecture leveraging a 3D Ferroelectric NAND (FeNAND) structure, providing significantly higher density, faster speeds, and lower voltage requirements compared to traditional NAND flash. Despite its superior density, the NAND structure has not been widely utilized in ISP applications due to limited throughput associated with row-by-row reads from serially connected cells. To overcome these limitations, we integrate hyperdimensional computing (HDC), a brain-inspired paradigm that enables highly parallel processing with simple operations and strong error tolerance. By combining HDC with the proposed dual-bound approximate matching (D-BAM) distance metric, tailored to the FeNAND structure, we parallelize vector computations to enable efficient MS spectral library search, achieving 43× speedup and 21× higher energy efficiency over state-of-the-art 3D NAND methods, while maintaining comparable accuracy.
\end{abstract}
 

\vspace{-2mm}

\section{Introduction}

 Mass spectrometry (MS) is crucial in proteomics and metabolomics, providing high-resolution analysis to detect molecular differences for drug development, personalized healthcare, and advancing biomarker discovery and disease pathway understanding. The rapid evolution of MS technology has led to vast data growth, with repositories like PRIDE \cite{martens2005pride} and MassIVE \cite{massive2024} exceeding 650 TB, creating challenges for efficient library searches essential to drug discovery and biomarker identification \cite{Duarte2016}. Traditional methods match experimental spectra to reference libraries but often fail to account for novel peptides or post-translational modifications (PTMs), leaving many spectra unannotated. Open modification searching (OMS) \cite{annsolo} matches MS/MS spectra irrespective of precursor mass—allowing detection of both modified and unmodified peptides despite PTM-induced mass shifts and decouples precursor mass from the search criteria to uncover a broader set of peptide‑spectrum matches and substantially improve PTM analysis. However, OMS significantly increases the search space, heightening computational and memory demands, which are further strained by the rapid growth of MS data and experiments, pushing current tools to their limits.

Modern MS analysis tools, whether based on conventional CPU or GPU architectures, often dedicate over 60\% of processing time to  matrix operations that require substantial memory\cite{hyperoms,homstc, Xu2023HyperSpec,pinge2023spechd}. 
Previous GPU-based approaches \cite{hyperoms,homstc}, leverage Hyperdimensional Computing (HDC) to accelerate OMS, benefiting from HDC’s parallel processing via simple binary operations, which is ideal for energy-efficient pattern recognition and associative memory tasks. However, HDC’s reliance on long vectors (1k–10k dimensions) exacerbates GPU memory limitations and data transfer costs, increasing latency and power inefficiency in large-scale OMS applications.

\begin{figure}[ht]
    \centering
    \includegraphics[width=0.43\textwidth]{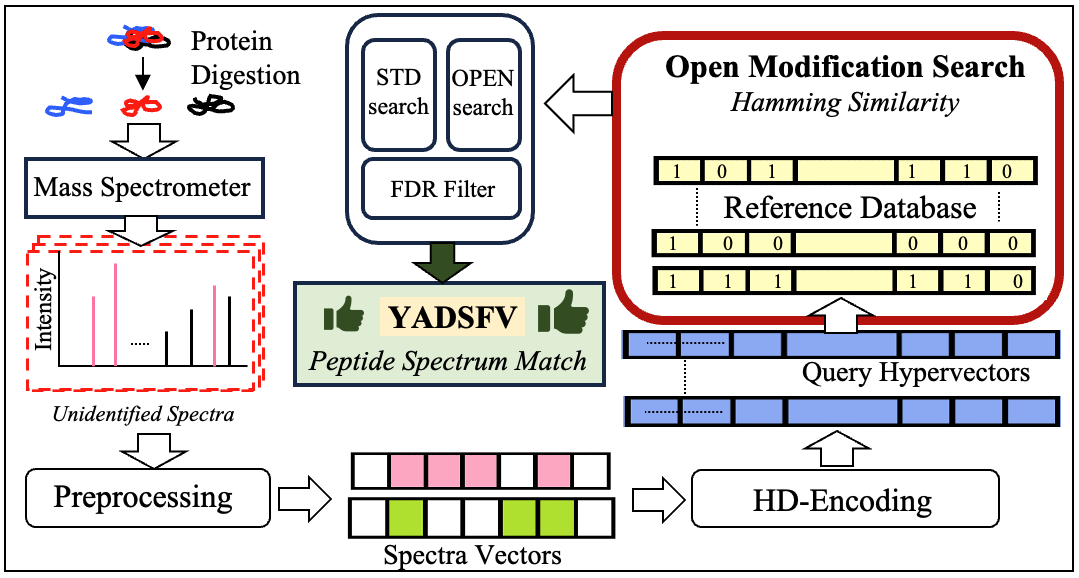}
    \caption{Open modification search using HDC.}
    \label{fig:pipeline}
\end{figure}

To overcome these challenges, previous work has employed processing-in-memory (PIM) to accelerate OMS with HDC, using memory technologies such as  RRAM, PCM, and NAND flash\cite{RRAM,Fan2024SpecPCM,ISP,Ponzina2025SmartMS}.
While PCM and RRAM offer advantages such as fast read operations and parallel energy-efficient matrix multiplication through analog computing, they face challenges such as lower density and limited scalability compared to NAND Flash, restricting their ability to meet the large data storage requirements of MS. Conventional 3D NAND based on charge-trap layers is nearing its density limits, but the integration of ferroelectric layers offers a promising pathway to exceed 100 Gb/mm$^2$ while reducing operating voltages and power consumption. Building on this, 3D FeNAND leverages ferroelectric layers to enhance scalability and performance, positioning it as a next-generation non-volatile memory solution.

Despite these benefits, NAND-based in-storage processing (ISP) faces utilization constraints due to serially connected structures, limiting ISP operations to row-by-row processing instead of simultaneous multi-wordline activation. Alternative approaches, such as voltage accumulation-based ISP in 2D NAND Flash, require extensive modifications, e.g., embedded resistive components\cite{resistive}, which are impractical for 3D FeNAND, increasing costs and reducing the storage density. 

We solve these challenges by introducing FeNOMS, a high-throughput, high-density ISP system leveraging 3D FeNAND technology to accelerate HDC-based mass spectrometry OMS search. The key contributions of this work are as follows:
 
\begin{enumerate}  
\item HD computing is utilized to maximize parallelism and simplify the computing kernel while ensuring error resiliency to tolerate non-idealities in ISP computations.
\item A novel distance metric, \textit{Dual-Bound Approximate Matching (D-BAM)}, is proposed to maximize throughput, by enabling computations directly within the FeNAND string and allowing simultaneous access to multiple rows with a configurable distance margin to tolerate both hardware and algorithmic non-idealities.
\item An FeNAND-based ISP architecture, with an external accumulator is proposed to process the distance calculation considering physical restrictions such as density, area, pitch, and noise while minimizing modification to be compatible to widely-used 3D NAND structure.
\item System and circuit-level validations on large-scale OMS datasets demonstrate accuracy and efficiency, achieving a 43× speedup and 21× higher energy efficiency over state-of-the-art 3D NAND architectures while addressing hardware non-idealities.

\end{enumerate}

\section{Background and Motivation}

\subsection{MS Pipeline \& OMS Search}\label{mspipeline}

 The MS flow (Fig. \ref{fig:pipeline}) begins with ionization, where molecules in a sample are charged and sorted by their mass-to-charge (m/z) ratios in a mass analyzer. The generated spectra are digitized, refined, and preprocessed to isolate key peaks, reduce background noise, normalize intensities, and bin the spectrum into extremely sparse vectors, steps that improve search quality yet add to the memory burden \cite{msas,hyperoms,Xu2023HyperSpec}. The processed data is searched against spectral libraries using various MS tools, with OMS \cite{annsolo} improving sensitivity by broadening the precursor m/z range and enabling OMS-specific detection of spectral variations \cite{hyperoms,homstc,Kong2017}. As illustrated in Fig.\ref{fig:pipeline}, OMS calculates the distance between an input query and each reference spectrum in a database. The top-$k$ reference spectra with the highest similarity scores are selected as candidates for the peptide-spectrum match. The primary computational bottleneck in OMS lies precisely in the distance calculations, which involve large-scale matrix operations and dominate runtime, as highlighted in studies \cite{Fan2024SpecPCM, homstc, pinge2024rapidoms}. This challenge is more noticeable on GPUs, especially when the dataset size exceeds the GPU onboard memory, leading to costly data movement overheads. To address this, we employ ISP, performing computations directly within memory to minimize data transfer costs and enhance efficiency.
 
\subsection{HD Computing for OMS}
\label{sec:background-hdc}
 Hyperdimensional computing (HDC) is a biologically inspired approach that encodes information in high-dimensional hypervectors (HVs) to enable resilient and parallelizable data representations\cite{Kanerva2009}. The processing stages are as follows:
 
\niparagraph{Bundling.}
Element-wise addition combines multiple HVs into one composite vector; for example, \( \mathbf{H}=A_1 + A_2 + A_3 \) remains similar to each constituent (\( d(\mathbf{H},A_i)\!\approx\!1 \)) yet nearly orthogonal to any unrelated random HV \( X \) (\( d(\mathbf{H},X)\!\simeq\!0 \)).

\niparagraph{Binding.}
Coordinate-wise multiplication (\(\circledast\)) associates different pieces of information: the product \( A\circledast B \) forms a HV that is nearly orthogonal to both operands (\( d(A\!\circledast\!B,A)\!\simeq\!0 \)), enabling compositional representation without losing randomness.

\niparagraph{HDC encoding.} Following established methods\cite{imani2017voicehd}, we employ an ID-level encoding approach to transform m/z and intensity values into a unified HD vector\cite{Xu2023HyperSpec}. This encoding uses two sets of binary hypervectors: ID HVs \( \{ I_1, I_2, \ldots, I_f \} \) to represent m/z values and level HVs \( \{ L_1, L_2, \ldots, L_Q \} \) to encode intensity levels. Each pair of m/z and intensity values \((i, j)\) is mapped to its corresponding ID HV \(I_i\) and level HV \(L_j\), which are then combined using a bitwise XOR operation \(I_i \oplus L_j\). Finally, a majority function is applied across the resulting binary values from all pairs \( (i, j) \), producing a binary hypervector \( \mathbf{h} \):

\begin{equation}
    h_i = \text{Majority}(L_{i1} \oplus I_{i1} + \cdots + L_{iD} \oplus I_{iD}). \tag{1}
\end{equation}

\niparagraph{HDC similarity check.}
HDC compares HVs by similarity metrics, such as Hamming distance for binary HVs or cosine similarity for non-binary HVs. Two HVs, \( \mathbf{H_1} \) and \( \mathbf{H_2} \), are considered distinct if \( d(\mathbf{H_1}, \mathbf{H_2}) \approx 0 \), which supports error-resilient retrieval and classification. For classification tasks, the class vector with the highest similarity score to the query is selected as the result. The similarity metric, extensively used in OMS workflows, often becomes a computational bottleneck, motivating our approach to address this challenge efficiently.

\subsection{In-storage processing within 3D NAND architectures} \label{sec:background-isp}
In-storage processing (ISP) in 3D NAND architectures enables efficient handling of large datasets by performing computations directly within the storage medium. Unlike traditional SSDs, where frequent data transfers between storage and processors cause significant latency and energy costs, prior 3D NAND ISP works applied input queries/vectors as voltages on bitlines (BLs) while storing reference data, such as weight matrices, in the NAND Flash. This allows operations such as matrix-vector multiplication (MVM) to occur seamlessly within the storage itself, reducing bottlenecks \cite{ISP}.

\begin{figure}[h]
\centering
\includegraphics[width=0.48\textwidth]{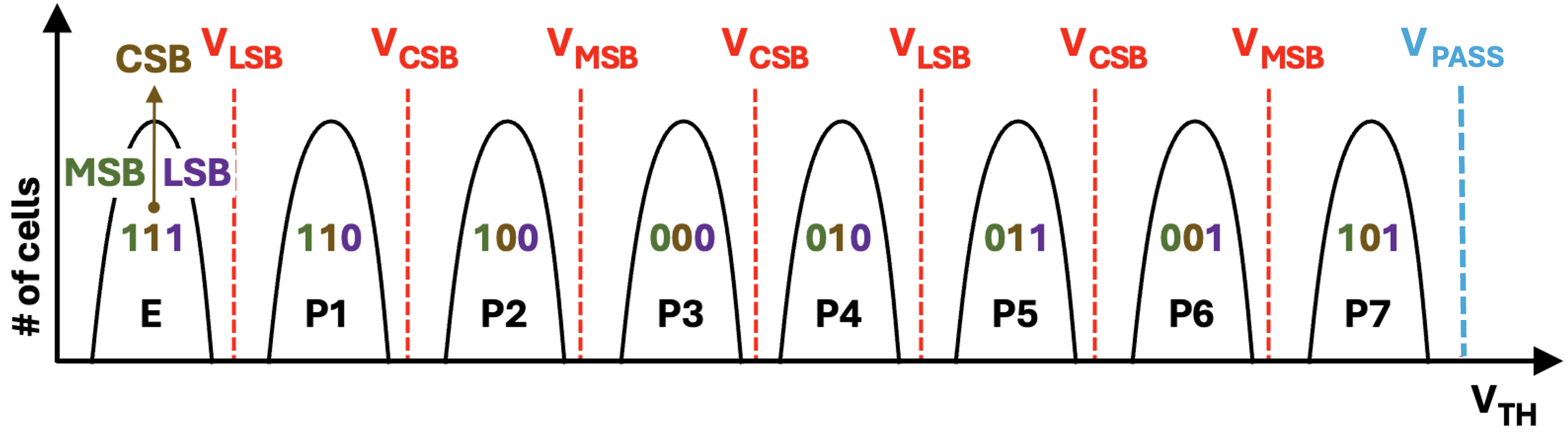} 
\caption{Seven sequential \(V_{\text{read}}\) sensing, for precise reading.}
\label{fig:precisereading}
\end{figure}

NAND flash memory in these architectures includes single-level cell (SLC) and multi-level cell (MLC) types, with MLC supporting variants like triple-level cell (TLC) and beyond. MLC stores data by programming cells with eight threshold voltage levels (\( V_{\textrm{Th}} \)), as shown in Fig.~\ref{fig:precisereading}, which TLC NAND distinguishes through seven wordline (WL) reads, requiring iterative sensing cycles. However, device variability from process variations and repeated program/erase (P/E) cycles causes shifts in \( V_{TH} \) distributions over time, impacting sensing accuracy and application reliability. Robust algorithms are critical to mitigate these effects and ensure reliable performance. 3D NAND ISP architectures face utilization constraints, allowing only one layer per block to be active at a time, significantly limiting cell access. Combined with iterative reads in Fig.~\ref{fig:precisereading}, this layer-by-layer reading further limits performance \cite{128,ISP}. While multiple WL activation is proposed in a few prior works in NAND\cite{choi2020binarynn,park2022flashcosmos,seshadri2017ambit,seshadri2015fast}, these techniques have been limited to SLC NAND or a few operands and are highly sensitive to non-idealities, making them unsuitable for similarity-based workloads requiring flexible matching. Despite these challenges, 3D NAND ISP provides significant gains in storage density and computational efficiency, making it ideal for in-situ processing on large datasets.


\subsection{High-Density, Energy-Efficient Ferroelectric NAND}
Over the past decade, 3D NAND has scaled from tens of layers to close to 300 layers. However, z-scaling in 3D NAND is limited by reliability challenges due to cell-to-cell interference and high write voltages \cite{han2023fundamental}. FeNAND has emerged as a compelling alternative to traditional Charge Trap Nitride (CTN)-based 3D NAND, addressing key limitations in z-scaling and data density \cite{lim2023comprehensive,das2023experimental,venkatesan2024disturb,han2023fundamental}. In FeNAND, the CTN layer is replaced with a ferroelectric layer having a spontaneous polarization and enables lower write voltages and provides a pathway for continued z-scaling. By supporting over 1000 layers and thus packing more storage capacity into the same physical area than CTN NAND, FeNAND achieves a breakthrough in vertical scaling that promises significantly higher capacities per chip, addressing the escalating data demands of AI, machine learning, and large-scale scientific simulations. By reducing the need for high programming voltages and eliminating charge migration, FeNAND enables lower power consumption, enhanced retention, and improved data integrity in addition to the higher data density. Moreover, FeNAND exhibits excellent endurance and rapid polarization switching, allowing frequent in-storage processing operations with minimal device wear over extended cycles. Such high-density architecture is particularly advantageous for parallel computing tasks, enabling ISP operations at scale and providing energy-efficient, high-capacity storage for data-intensive applications.\cite{Lee2024fundamental}

\begin{figure*}[!ht]
\centering
\includegraphics[width=1\textwidth]{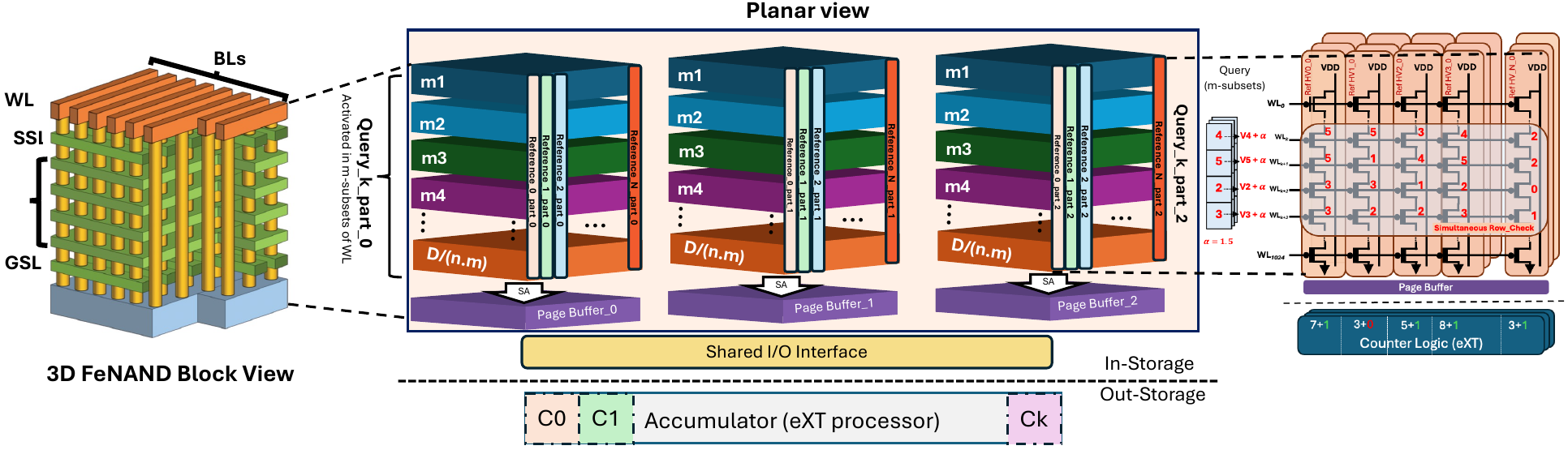}
\caption{FeNOMS system overview with data mapping on 3D FeNAND array and dataflow for D-BAM implementation.}
\label{fig:overview}
\end{figure*}


\section{Proposed FeNOMS ISP Acceleration Framework}

When dealing with large-scale data applications like OMS, the choice of memory architecture is a critical factor. Among various non-volatile memory technologies, 3D FeNAND stands out due to its significantly higher memory density compared to alternatives like ReRAM and PCM. This high density makes 3D FeNAND particularly well-suited for implementing large-scale in-storage computations. Unlike other technologies that might require frequent reloading of data to accommodate all data points, 3D FeNAND's capacity enables seamless handling of extensive datasets within a single memory structure, reducing overhead and improving computational efficiency.

While appealing, the 3D FeNAND structure suffers from significant read latency issues, as it requires sequential activation of the wordlines to retrieve a full word from the string. Moreover, detecting the multi-bit values stored in MLC necessitates multiple read operations to correctly determine the stored data, as elaborated in Section~\ref{sec:background-isp}. To address the high read latency problem, we propose FeNOMS, a novel ISP framework optimized for distance check operations, tailored specifically to the 3D structure of FeNAND memories. Our approach focuses on leveraging and enhancing the existing capabilities of 3D FeNAND while minimizing the need for extensive modifications to the circuitry. By relying on the infrastructure already present in commodity memory chips, our design ensures practical implementation and broader applicability in real-world systems.

\subsection{FeNOMS ISP Overview} \label{sysoverview}

\niparagraph{System overview.} Our proposed system, illustrated in Fig.~\ref{fig:overview}, integrates a 3D FeNAND memory for storing reference HVs of the database and performing ISP-based distance checks between query and stored references with a lightweight external processor for completing the post-processing stages.
To ensure scalability, the architecture of the 3D FeNAND and its peripherals closely follows conventional industrial structures, making it practical for real-world implementation. While most of the distance computations are processed internally in 3D FeNAND in a highly parallel manner, minor post-processing operations are performed on the accumulator (external processor).
The ISP-based distance checks in 3D FeNAND follow our proposed Dual-Bound Approximate Matching technique, which minimizes the data movement between the memory unit and the external processor. 
The post-processing operations, handled by the processor, include the accumulation of the ISP output data and False Discovery Rate (FDR) filtering, contributing a minimal fraction to the total runtime\cite{Fan2024SpecPCM}.

\begin{figure}[h]
\centering
\includegraphics[width=0.48\textwidth]{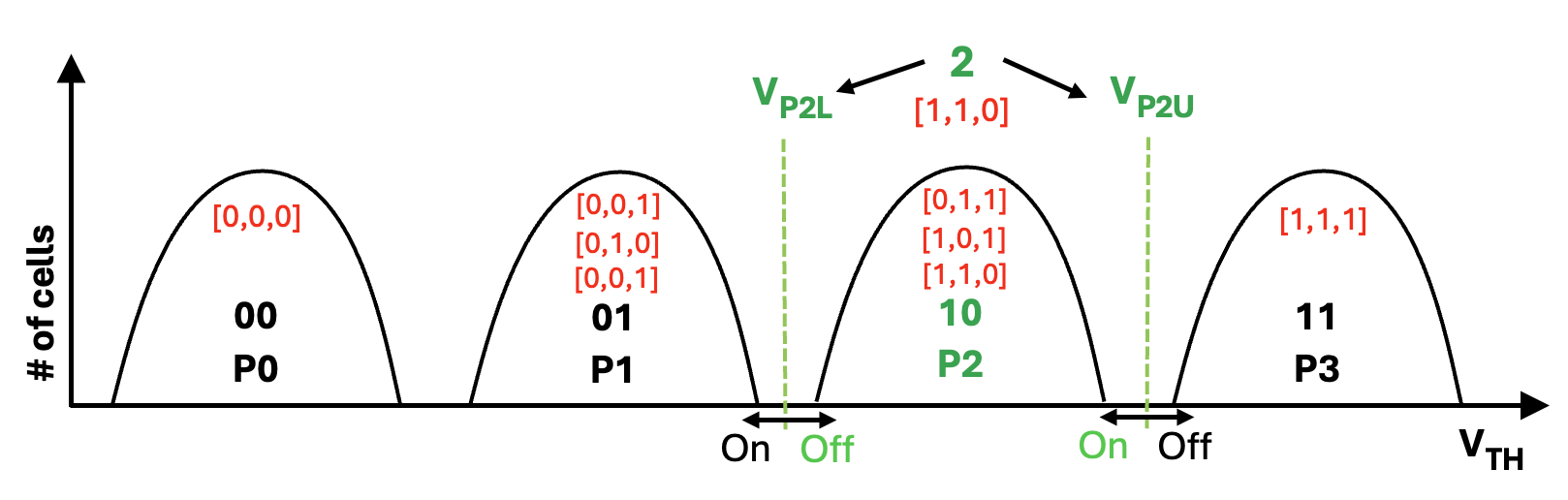} 
\caption{Sensing with \( \text{PF}_3 \), optimized for similarity search.}
\label{fig:similaritysearch}
\end{figure}

\niparagraph{Pre-processing stage.} HVs are generated during a pre-processing stage using the HDC encoding algorithm detailed in Section~\ref{mspipeline}. The encoded reference HVs are stored in FeNAND strings using a dimensional packing method, leveraging the multi-level programming capabilities of 3D FeNAND. Dimensional packing compresses a binary vector of length \( D \) into a shorter vector of length \( D/\text{PF}_n \) by summing \( \text{PF}_n \) (packing factor) adjacent binary bits into an integer value, where \( \text{PF}_n \) represents the number of adjacent bits considered. The bits stored per MLC depend on \( \text{PF}_n \)  (Fig. \ref{fig:similaritysearch}); e.g., \( \text{PF}_3 \) yields 2 bits, while \( \text{PF}_4 \) or \( \text{PF}_5 \) yield 3 bits. This approach aligns with the standard practice of storing reference data once and reusing it multiple times without regeneration, as demonstrated in previous works \cite{pinge2024rapidoms,hyperoms}, resulting in minimal overhead.

\niparagraph{HD data mapping on 3D FeNAND.} Fig.~\ref{fig:overview} further illustrates the detailed data mapping within the memory array, which is organized into multiple planes, each comprising multiple 3D FeNAND blocks. Given the high dimensionality of HVs (e.g. 8K bits) and the layer limitations of FeNAND (e.g., up to 512 wordlines per string \cite{512}), HVs are folded and distributed across vertical strings located on different blocks within a plane. 
For distance computation, the query HV is streamed through the wordlines, enabling interaction with the stored reference HVs within each layer. Similar to the storage process, query HVs exceeding the FeNAND's layer capacity are also folded into multiple pages, with each portion of the query compared against the corresponding portion of the references as previously distributed.
\niparagraph{Parallel ISP-based distance calculation.} Unlike conventional NAND-based architectures, we enable multiple ($m$) wordlines to access multiple layers simultaneously, thereby parallelizing the distance search process for higher throughput. This approach requires only minor modifications to the wordline decoder in FeNAND, resulting in negligible overhead. While it facilitates parallel comparisons of multiple elements and enhances throughput, it also introduces potential accuracy degradation, as simultaneous comparison of multiple values increases the likelihood of mis-comparisons due to the inability to evaluate each element individually. To address these issues, we propose the Dual-Bound Approximate Matching (D-BAM) which is described in the following section.  The effect of varying the number of parallel layers is analyzed in Section~\ref{results:accuracy}.

\subsection{Dual-Bound Approximate Matching (D-BAM)}\label{sec:dbam}

\begin{figure}[h]
    \centering
    \begin{subfigure}[t]{0.22\textwidth} 
        \centering
        \includegraphics[width=\textwidth]{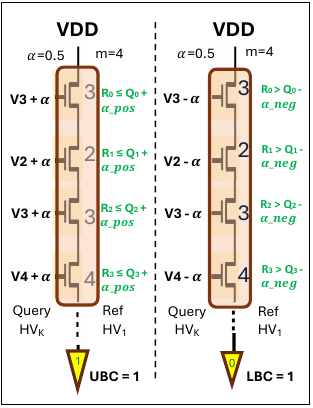}
        \caption{Match case}
        \label{fig:match_diagram}
    \end{subfigure}
    \hfill
    \begin{subfigure}[t]{0.224\textwidth} 
        \centering
        \includegraphics[width=\textwidth]{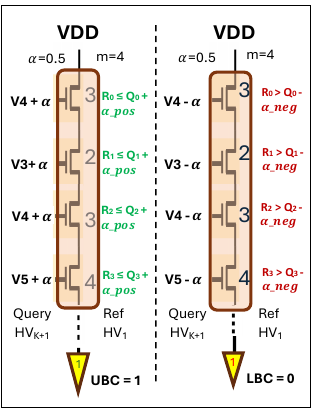}
        \caption{Unmatch case}
        \label{fig:unmatch_diagram}
    \end{subfigure}

    \vspace{0.20cm} 

    \begin{subfigure}[t]{0.24\textwidth} 
        \centering
        \includegraphics[width=\textwidth]{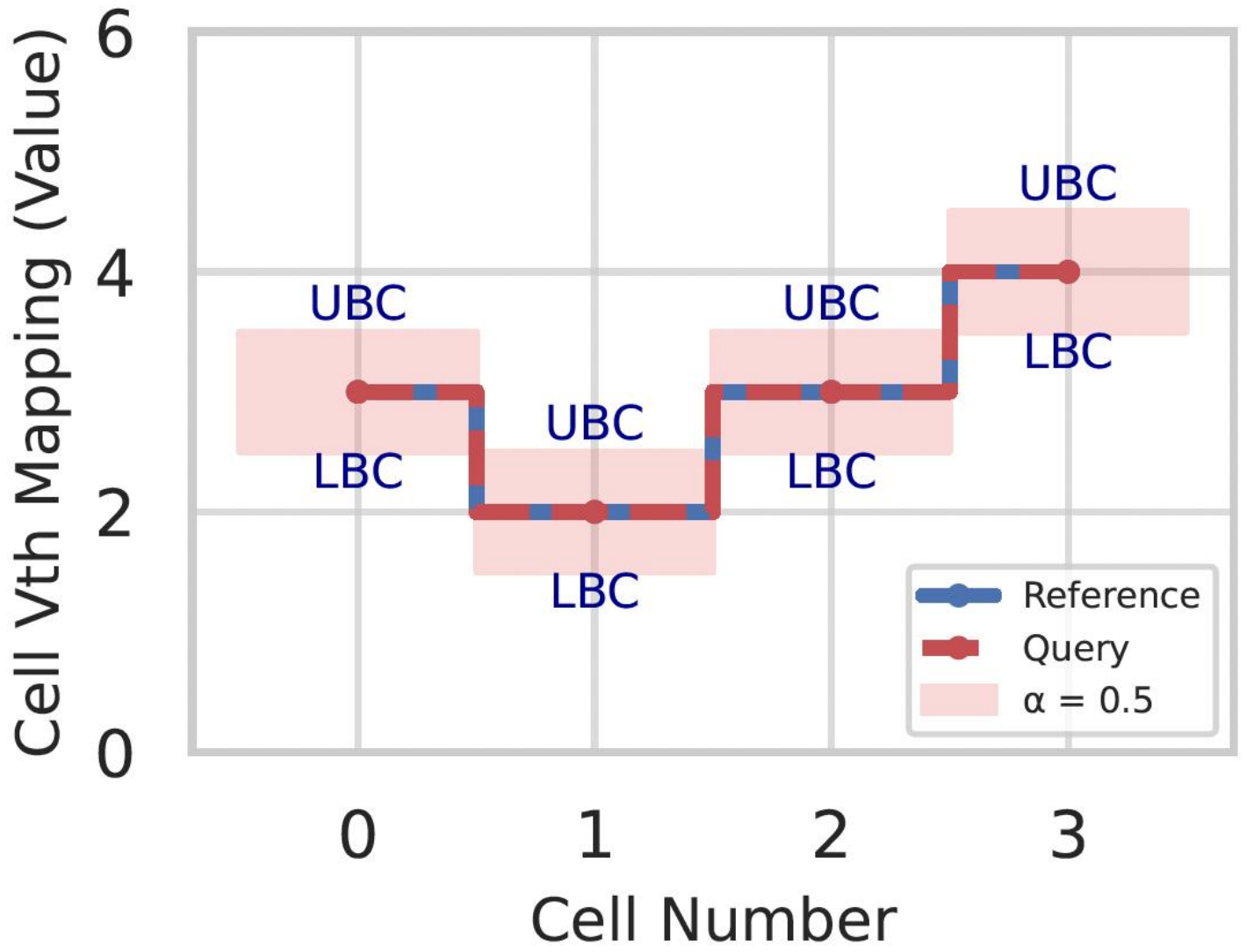}
        \caption{Match: D-BAM mapping.}
        \label{fig:match_ubc}
    \end{subfigure}
    \hfill
    \begin{subfigure}[t]{0.24\textwidth} 
        \centering
        \includegraphics[width=\textwidth]{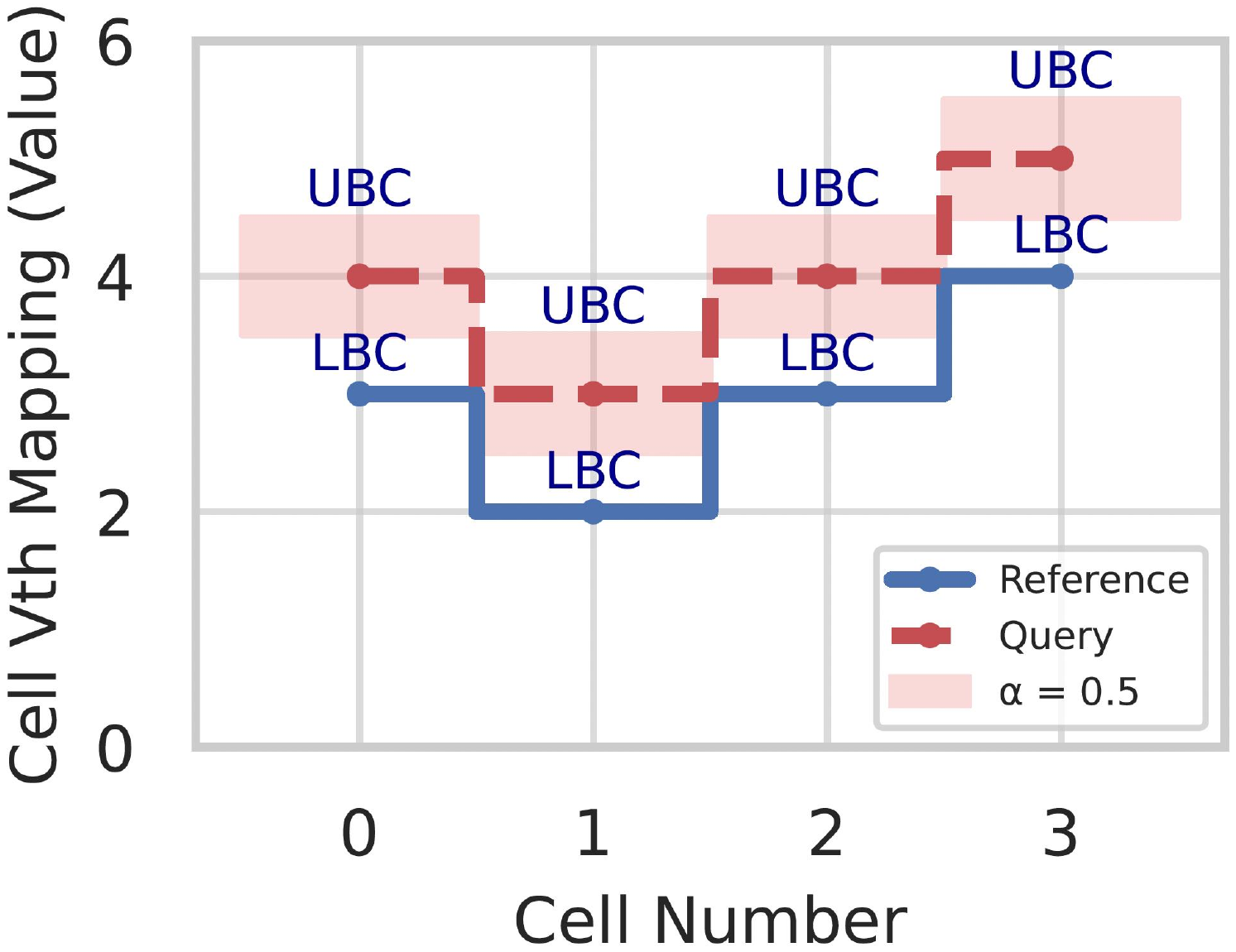}
        \caption{Unmatch: D-BAM mapping.}
        \label{fig:unmatch_ubc}
    \end{subfigure}
\caption{D-BAM tolerance-based similarity checks: query \( q \) as wordline voltages, reference \( V_{\text{TH}} \) in FeNAND string.}
    \label{fig:quadrant}
\end{figure}

 Dimension-packing, previously applied in a prior work~\cite{Fan2024SpecPCM}, cannot support matrix-vector multiplication with multiple-row activation in 3D FeNAND arrays due to FeNAND cell's high on/off ratio (around $10^8$ \cite{Si2019}). Dimension-packing combines multiple binary bits into single integer values, which may increase mismatch likelihood. However, approximate matching is effective for OMS. Non-ideality in device characteristics—such as threshold voltage shifts from process variation, wear, and environmental factors such as temperature and voltage drop can further contribute to mismatches. To address these issues, we propose tolerance-based similarity checks: D-BAM, which follows dimension-packing to enable accurate comparison between reference and query vectors with a tolerance margin ($\alpha$), as illustrated in Fig.~\ref{fig:quadrant}. This approach performs two checks as described below.

\begin{itemize}
    \item \textbf{Upper Bound Check (UBC):} A UBC  is conducted first by setting the wordline voltage to $q_i + \alpha_{\text{pos}}$ where $q$ and $r$ are query and reference vector elements, respectively. By applying a voltage slightly above the target threshold, we assess if $r_i$ falls  below this upper limit. Current flow through the FeNAND string at this voltage indicates that $r_i \leq q_i + \alpha_{\text{pos}}$, passing the UBC check. For the $m$ elements parallel comparison, the current flows only when all the elements are below the upper bound (UB). This is formalized as:
    \begin{equation}
    \text{UBC}_j = \prod_{i=mj}^{mj+m-1} \left[ r_i \leq q_i + \alpha_{\text{pos}} \right]
    \end{equation}
    If no current is detected, the positive check fails, indicating at least one value in the $m$-subset exceeds the tolerance.

    \item \textbf{Lower Bound Check (LBC):} Following the UBC, an LBC  is performed by setting the wordline voltage to $q_i - \alpha_{\text{neg}}$, which is a lower bound (LB). In this case,  $r_i$ values  should be above the LB and the lack of current flow confirms that $r_i \geq q_i - \alpha_{\text{neg}}$, passing the negative check. By performing LBC for $m$ parallel elements, no current in the serially-connected string  means at least one element has passed the LB check. Unlike UBC, which requires all elements to meet the condition to pass, LBC employs a more lenient approach, passing if any element satisfies the condition. While this method is suboptimal in terms of accuracy, it offers the advantage of requiring no modifications to the array, making it seamlessly compatible with serially connected strings. An LBC value of ``1" indicates that the check has passed the check as formalized in the following equation:
    

\begin{equation}
\text{LBC}_j = 1 - \prod_{i=mj}^{mj+m-1} \left[ r_i < q_i - \alpha_{\text{neg}} \right]
\end{equation}
 \end{itemize}

These checks collectively ensure that the reference values are within the allowable tolerance bounds for all elements in the $m$-subset. Scores are then updated for each \( m \)-subset based on the UBC and LBC results across all dimensions as follows, where $n$ shows the number of binary values in dimension packing and D is the dimension of hypervectors.


\begin{equation}
\text{Score} = \sum_{j=1}^{\frac{D}{mn}} \text{UBC}_j + \sum_{j=1}^{\frac{D}{mn}} \text{LBC}_j
\label{score}
\end{equation}
As scores are accumulated across multiple UBC and LBC subsets, this approach ensures a high-quality similarity check with strong resilience, even when a specific subset fails to match due to partial mismatches or hardware-induced non-idealities.

The \(\prod\) function is inherently suited to the FeNAND string due to its serial connectivity. Notably, we seamlessly embedded the distance metric into the hardware without altering any array structure and read/write circuitry, and both checks can be performed using a single sense amplifier (SA). 
Enabling multiple rows simultaneously may increase the effective resistance due to $m$ cascaded turned-on cells. However, the on/off ratio of $10^8$\cite{Si2019} in FeNAND provides a sufficient margin to reliably distinguish the current flow.
By leveraging the above similarity checks with some margin ($\alpha$), this approach effectively assesses similarity between high-dimensional vectors, tolerating both dim packing-induced discrepancies and hardware non-idealities. The effect of $\alpha$ on the search quality is analyzed in Section~\ref{results:accuracy}. 


Employing D-BAM significantly accelerates similarity searches, not only due to the $m$-parallel comparisons but also because it requires only a single read access (comparison). In contrast, conventional MLC approaches necessitate iterative read operations, as depicted in Fig.~\ref{fig:precisereading}, by incrementally adjusting the wordline voltages. Consequently, assuming $n$ binary values are dimensionally packed, $2^{n-1}$ read accesses are required in the conventional scenario. This substantial reduction in read operations leads to a remarkable speedup factor achieved by D-BAM compared to the conventional row-by-row reading approach, which can be expressed as:

\begin{equation}
\text{Speedup} = \frac{\text{MLC}_n}{\text{PF}_n \, (\text{with D-BAM})} = \frac{m \cdot (2^n - 1)}{2}
\end{equation}

The factor \( 2 \) in the denominator accounts for the fact that both UBC and LBC checks are required. As an example, for D-BAM with \( m = 4 \), it results in a \( 14\times \) speedup in the read operations for TLC (\( n = 3 \)) and a \( 30\times \) speedup for QLC (\( n = 4 \)). 

\niparagraph{D-BAM support in 3D FeNAND and hardware overhead.} 
The UBC and LBC operations require detecting current flow through the bitlines and forwarding the results to the column decoder in a time-multiplexed manner, leveraging the existing conventional decoding logic, sense amplifiers, and I/O path.  The sense amplifier output, which provides the condensed single-bit results of UBC and LBC, is then directly delivered to the page buffer, which subsequently transfers the results off-chip.
It is important to note that a conventional MLC-based NAND typically requires a multi-bit page buffer for each string. In contrast, D-BAM produces only binary outputs. As a result, D-BAM can be implemented without requiring any additional buffers or sense amplifiers, instead reusing a subset of the existing page buffer to deliver merely binary bit per string through the existing I/O path. This enables full compatibility with the existing page buffer architecture.

The bits from the page buffer are then relayed to an external processor for score accumulation, as described in (\ref{score}), using simple binary counters. While one can consider integrating the counter logic within the memory module near the page buffer, doing so would result in significant area overhead due to the highly compact structure of FeNAND strings. As a result, the binary counter is implemented outside the memory module to maintain memory density and the conventional interface. 

The only hardware addition required to support D-BAM is an extended wordline decoder that enables the activation of multiple consecutive wordlines through minimal logic modifications—for example, adding a few OR gates and control signals to indicate the number of lines to be enabled. 
Importantly, the  proposed modifications introduce negligible overhead—$<$0.5\%—in area  and energy consumption, as the decoder logic is shared across all columns~\cite{gonugondla2018energy}.



\section{RESULTS}

\subsection{Experimental Setup}

\niparagraph{Datasets and quality metrics.} We evaluate our design using a real-world, large-scale MS dataset. Reference datasets include the human spectral library with the HEK293 dataset (b1906-b1932)\cite{hek} as the query. For detailed analysis, such as the UpSet plot, a subset of HEK293 (b1906) was used to represent a consensus. Quality is holistically assessed based on the number of identifications and their overlap with the consensus \cite{consensus,homstc}.

\niparagraph{Hardware configurations and peripheral circuit summary.} 
FeNAND-based energy and speed are modeled using the 3D NAND architecture \cite{ISP, SLC3DNAND} as a baseline, incorporating a z-dimension scaling factor  to account for shorter string lengths in FeNAND  compared to conventional 3D NAND. This scaling factor $k=4$, derived from in-house modeling and inference from previous studies on ferroelectric material properties \cite{das2023ferroelectric,512,lim2023comprehensive,das2023experimental,venkatesan2024disturb,han2023fundamental, kim2021ferroelectric}, is applied for the  RC delay modeling and analysis to compare with conventional 3D NAND (CTN).
The proposed FeNAND ISP architecture leverages heterogeneous integration to minimize area and power overhead~\cite{SLC3DNAND, techinsights2020, caillat2017, chen2020}. In this design, control circuits including the wordline/string select line switch matrix  and pass transistors are implemented using a CMOS under-array (CUA) approach at an effective 65~nm node. Other low-voltage digital peripheral circuits are derived using NeuroSim~\cite{NeuroSim} at a 1~GHz clock frequency and integrated with the 3D NAND array through face-to-face Cu--Cu hybrid bonding, employing an inter-tier pitch of 1~\(\mu\)m to ensure high-bandwidth communication. The accumulator unit in the external processor is designed in Verilog and synthesized with a 65~nm CMOS process using the Cadence Genus tool at a 1~GHz clock frequency. The integration strategy and modeling approach for these circuits are adopted from~\cite{SLC3DNAND}. FeNOMS employs a 3D FeNAND memory for an ISP capability  with the hardware configurations listed in Table~\ref{table:simparam}. Detailed evaluations of area, power, and implementation complexity are provided in 
Section~\ref{results:hardware}.

\begin{figure}[h]
\centering
\includegraphics[width=0.45\textwidth]{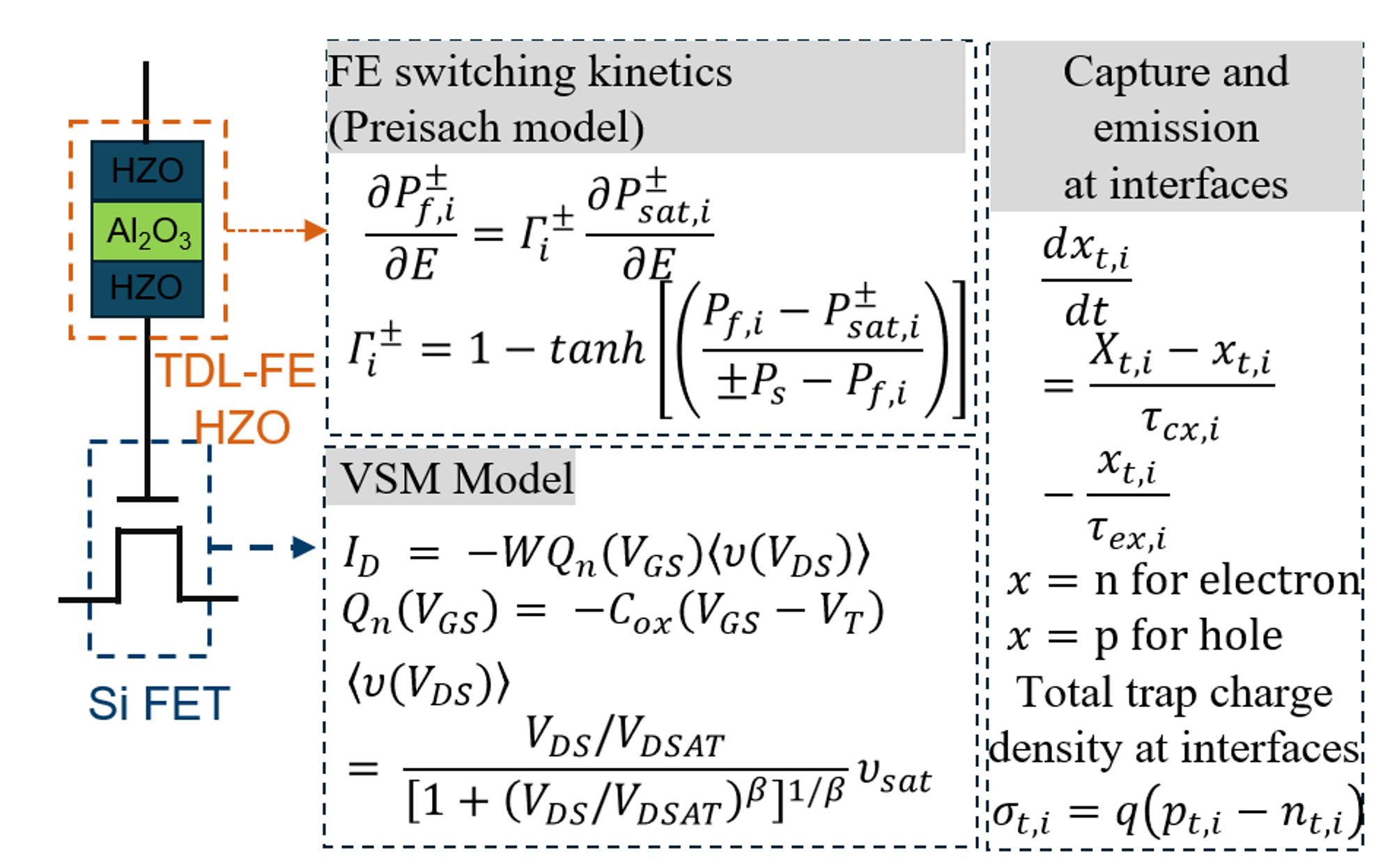} 
\caption{Trapping dynamics and switching kinetics in TDL FE with Virtual Source Model (VSM) for Si FET.}
\label{fig:feFET_overview}
\end{figure}

\begin{figure}[h]
\centering
\includegraphics[width=0.31\textwidth]{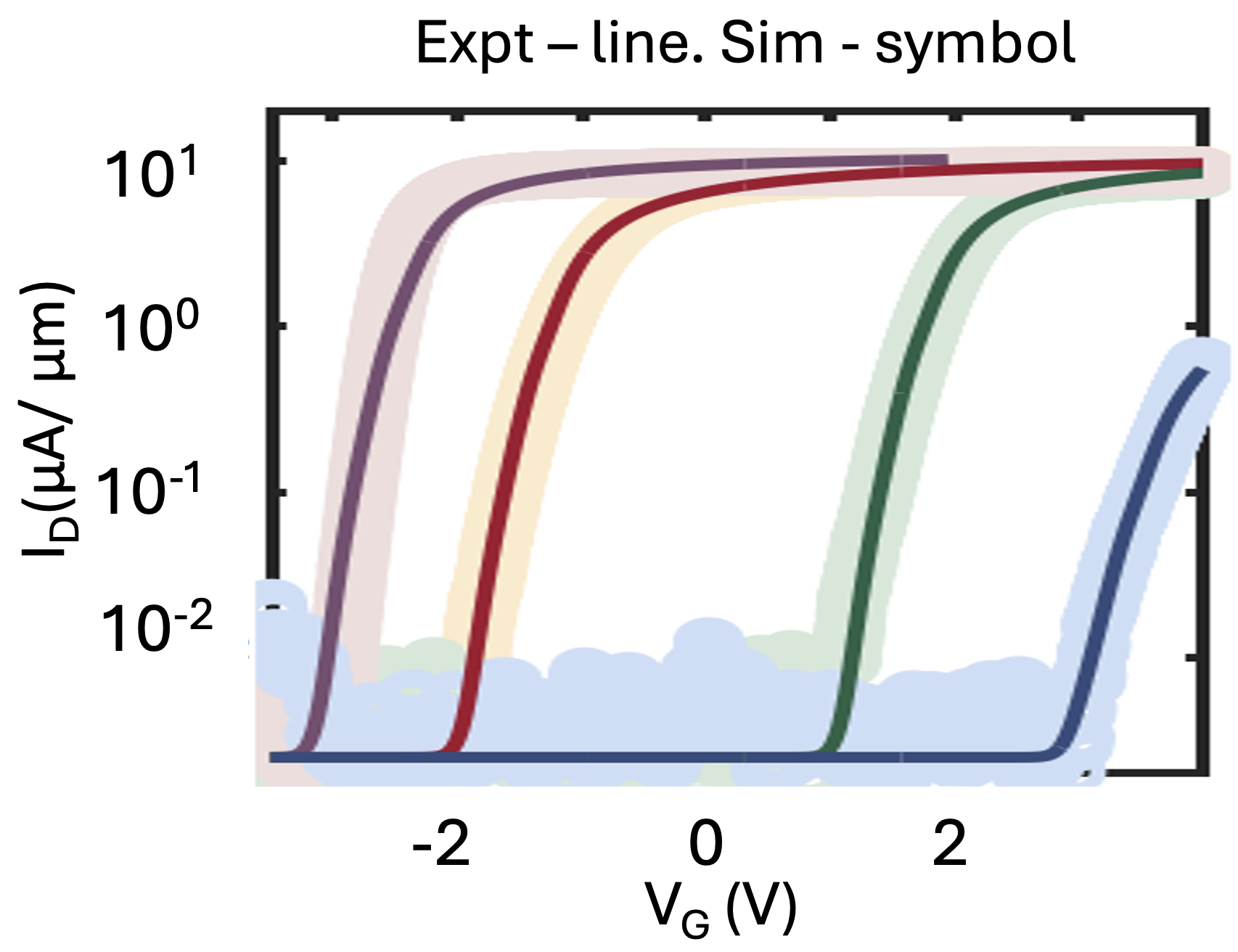} 
\caption{The experimental $I_{\mathrm{D}}$–$V_{\mathrm{G}}$ data from device simulations.}
\label{fig:feFET_experiments}
\end{figure}

\niparagraph{Device simulations and hardware non-idealities modeling.} Fig.~\ref{fig:feFET_overview} illustrates the modeling framework of a ferroelectric field‑effect transistor (FeFET) incorporating a ferroelectric stack based on HfZrO$_2$ (HZO) with an interfacial Al$_2$O$_3$ dielectric, integrated on top of a silicon FET. Ferroelectric polarization switching is described using a Preisach‑based kinetic model\cite{Wang2021bu}, where the polarization change is governed by a field‑dependent switching factor that captures the hysteretic nature of the ferroelectric material. The underlying FET operation is modeled using the Virtual Source Model (VSM). Additionally, the dynamics of charge trapping and de‑trapping at interfaces are modeled through rate equations, incorporating carrier‑type dependent capture and emission times. The total interfacial charge is then computed based on the trap occupancy, which plays a crucial role in modulating the threshold voltage and overall device hysteresis. 

Fig.~\ref{fig:feFET_experiments} shows the simulation of an FeNAND with a 6.5\,V memory window (MW), extracted from calibration against experimental $I_D$–$V_G$ curves. 
The device variation in terms of $\sigma_{V_T}$ is estimated as a function of device geometry based on Pelgrom’s law scaling ($\sigma^2 = A^2/2WL$)\cite{Kaczmarek2023}. Given our target parameters in Table~\ref{table:simparam}, the threshold‑voltage variation of $\sigma_{V_T}\approx200\,\text{mV}$ is obtained and used for our analysis. To capture these non‑idealities in our ISP simulations, Gaussian noise $N \sim \mathcal{N}(0,0.2^2\,\text{V})$ is applied for the noise-aware simulations, while enforcing the  6.5\,V memory window.

\niparagraph{Baseline designs.} To compare FeNOMS with previous work, we benchmarked it against state-of-the-art (SoTA) tools, including ANN-SoLo \cite{annsolo}, SpectraST \cite{Ma2014}, and HDC-based methods such as HOMS-TC\cite{homstc} and HyperOMS\cite{hyperoms}. ANN-SoLo and SpectraST were excluded from speed and energy efficiency comparisons due to their extremely slow performance, requiring more than 20× the runtime of other methods. HOMS-TC employs INT8 HV representation, HyperOMS uses a binary format, and FeNOMS leverages MLC configurations with bit-packing for efficient storage. Performance and energy efficiency metrics are based on simulation traces from HyperOMS software running on an NVIDIA GeForce RTX 4090 GPU. For fair comparison, HV dimensions (vector length) were kept consistent (8k) across FeNOMS, HOMS-TC, and HyperOMS. The evaluation also includes comparisons against a prior 3D NAND SLC-based OMS design and its modified version for MLC NAND using a similar ISP algorithm \cite{SLC3DNAND}.

\begin{table}[htbp]
\centering
\caption{Hardware Simulation Parameters Comparison.}
\renewcommand{\arraystretch}{1.3} 
\setlength{\tabcolsep}{6pt} 
 \scriptsize
\begin{tabular}{|l|c|c|}
\hline
\multicolumn{3}{|c|}{\textbf{3D FeNAND Simulation Parameters}} \\
\hline
\textbf{Parameter} & \textbf{SoTA Comparisons} & \textbf{FeNAND DSE} \\
                   & (WL = 32, \#Planes = 23)   & (WL = 512, \#Planes = 2) \\
\hline
SSL, WL, BL Pitch & 220 nm, 500 nm, 100 nm & 220 nm, 500 nm, 100 nm \\
\hline
FeNAND z-scaling & \multicolumn{2}{c|}{\(k = 4\)} \\
\hline
WL, SSL, Blocks & 32, 16, 128 & 512, 16, 128 \\
\hline
Config & SLC/TLC/PF\textsubscript{3}/PF\textsubscript{4} & PF\textsubscript{2}/PF\textsubscript{3}/PF\textsubscript{4} \\
BL & 16384, 5462, 5462, 4192 & 8192, 5462, 4096 \\
Plane area (mm\(^2\)) & 0.757, 0.252, 0.252, 0.189 & 0.378, 0.252, 0.189  \\
\hline
WL, SSL, BL Read & \multicolumn{2}{c|}{1 V, 4.5 V, 0.2 V} \\
\hline
\end{tabular}
\label{table:simparam}
\end{table}

\subsection{Search Quality Analysis}\label{results:accuracy}
\begin{figure}[h]
    \centering
    \includegraphics[width=0.495\textwidth]{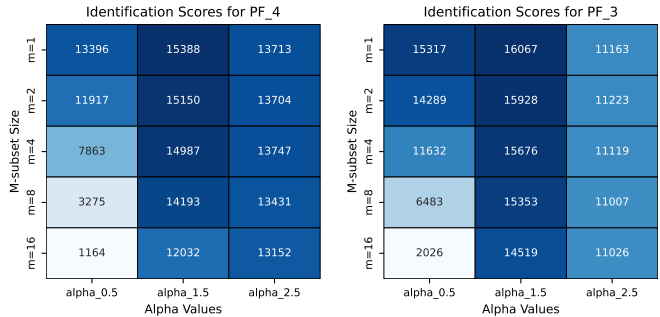}
    \caption{FeNOMS Identifications vs. tolerance ($\alpha$) and \( \text{PF}_n \). }
    \label{fig:heatmap}
\end{figure}

Fig. \ref{fig:heatmap} is the heatmap illustrating peptide identifications with FeNOMS across alpha scaling (\( \alpha=0.5, 1.5, 2.5 \)) and $m$-subset scaling (\( m=1, 2, 4, 8, 16 \)). \( \alpha \) = 0.5 corresponds to stricter thresholds near exact matches, resulting in fewer identifications. At \( \alpha = 1.5 \), identification rates remain robust with minimal drop-off up to \( m = 8 \). For larger \( m \), such as \( m = 16 \), higher \( \alpha \) values are needed to compensate for the inaccuracy from  increased parallelism, allowing relaxed thresholds to enhance tolerance. Notably, high similarity between query and reference does not directly correlate to identification. A very relaxed $\alpha = 2.5$ can lead to false hits during top-$k$ selection, where mismatched spectra may score highly, ultimately lowering identification rates. This underscores the importance of optimizing $\alpha$ and $m$ to balance tolerance and parallelism in peptide identification.

\begin{figure}[h]
    \centering
    \includegraphics[width=0.49\textwidth,page=1]{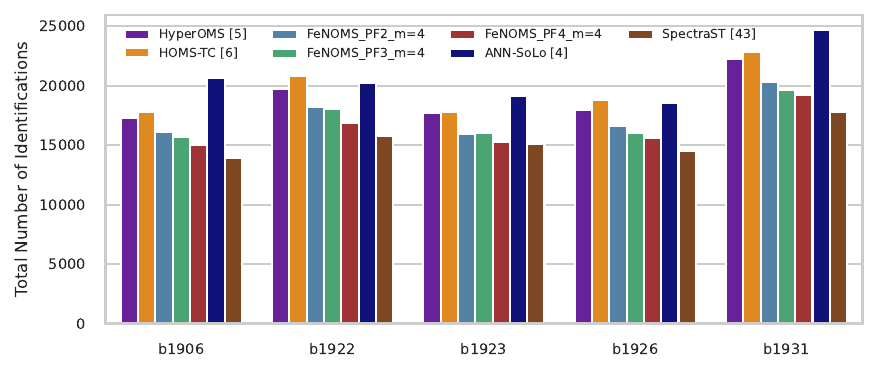}
    \caption{\#Identifications of FeNOMS vs. SoTA.}
    \label{fig:identifications_comparison}
\end{figure}

\begin{figure}[h]
    \centering
    \includegraphics[width=0.39\textwidth]{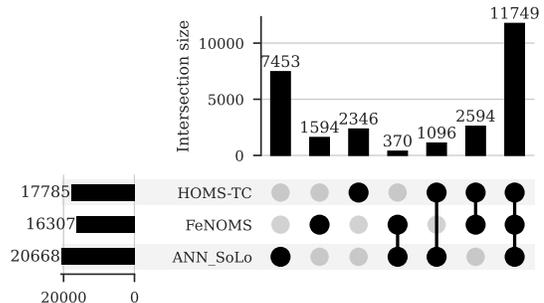} 
    \caption{Consensus UpSet plot: FeNOMS vs. SoTA}
    \label{fig:upset_plot}
\end{figure}

Fig. \ref{fig:identifications_comparison} compares the total number of peptides identified across different configurations with \( \alpha = 1.5 \) and $m=4$ for queries b1906–b1931.  It is important to understand that total identifications alone do not fully assess tool performance. As shown later in Fig. \ref{fig:upset_plot}, overlap analysis provides deeper insights into a tool's ability to capture high-confidence peptides.  FeNOMS outperforms SpectraST\cite{Ma2014} in peptide identification and shows comparable performance to HDC-based methods HOMS-TC (INT8) \cite{homstc} and HyperOMS\cite{hyperoms}. For similar HD algorithms, a drop in identification accuracy is generally expected due to the compression in MLC configurations and the use of $m$-subset checks. Despite this, FeNOMS demonstrates robust performance, maintaining computational efficiency with only a minor reduction in candidate identifications (e.g., FeNOMS (PF$_2$, $m=4$) yields approximately 6\% fewer candidates than baseline HyperOMS\cite{hyperoms}. Although ANN‑SoLo~\cite{annsolo} identifies the most peptides, its lengthy runtime and high
power consumption severely limit its practicality for large‑scale OMS.

\begin{figure}[h]
    \centering
    \includegraphics[width=0.49\textwidth,page=1]{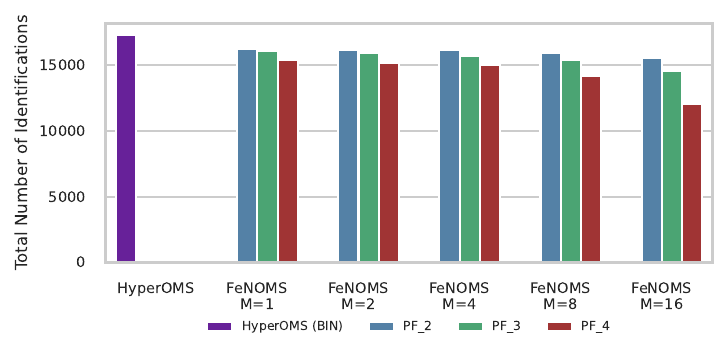}
\caption{Effect of \(PF_{n}\) and \(m\) on identifications.}
    \label{fig:row_scaling}
\end{figure}

Fig. \ref{fig:row_scaling} shows the impact of $m$ and $\textrm{PF}_n$  by using query b1906 which has an average number of peptide identifications. Binary SLC HyperOMS\cite{hyperoms} is included as the baseline. FeNOMS performs strongly at lower \( m \)-subset sizes, with minimal variation in identifications. Even at high parallelism (\(m = 8\)), FeNOMS retains over 90\% of baseline identifications, underscoring its robustness. Under extreme parallelism (\(m = 16\)), identification rates decline more sharply, particularly for the highest packing factor $\textrm{PF}_4$, highlighting the throughput–accuracy trade‑off. Overall, $\textrm{PF}_2$ and $\textrm{PF}_3$ configurations maintain high identification rates for up to \(m = 8\), providing robust performance with minimal degradation.

In the absence of ground truth for the HEK293 dataset, we evaluate identification accuracy performance based on the number of identified spectra in consensus, a well‑established proteomics metric~\cite{consensus,hyperoms,annsolo}. Fig.~\ref{fig:upset_plot} compares the overall search quality achieved by FeNOMS, HOMS‑TC~\cite{homstc}, and ANN‑SoLo~\cite{annsolo} using an UpSet plot. FeNOMS demonstrates strong identification capabilities, capturing a substantial number of peptides and showing significant overlap with established tools such as HOMS‑TC and ANN‑SoLo. Of the 12845 peptides in the HOMS‑TC \(\cap\) ANN‑SoLo consensus (11749 + 1096), FeNOMS recovers approximately 92\% of that joint set. This high overlap underscores the reliability of FeNOMS and confirms its consistency with current state‑of‑the‑art methods. Notably, peptides uniquely identified by FeNOMS (1,594 in total) may represent novel findings and boost overall coverage by  12\% without compromising the controlled false discovery rate, further highlighting FeNOMS’s ability to extend search sensitivity beyond existing approaches.

\subsection{PPA Improvement Analysis}\label{results:hardware}

We conduct two sets of analyses based on configurations in Table~\ref{table:simparam}: the first compares our results with SoTA, using 32 WLs, which was employed in \cite{SLC3DNAND}, where conventional 3D NAND struggles to scale to very large layers. The second focuses on FeNAND with 512 WLs, incorporating design space exploration under different configurations as shown in Fig.~\ref{fig:row_scaling}. For all analyses, we ensure the same storage capacity by scaling bitlines while keeping the number of blocks and planes constant, as blocks and planes provide parallelism scaling, enabling fair comparisons as detailed in Table~\ref{table:simparam}.

\begin{table}[ht]
\scriptsize
\centering
\caption{Comparison of PPA with SoTA.}
\begin{tabular}{|c|c|c|c|}
\hline
\textbf{Approach} & \textbf{Latency [s] } & \textbf{Energy [mJ]} & \textbf{Area [mm$^2$]} \\
    & (speedup×) & (efficiency×) & (efficiency×) \\ \hline
\textbf{HyperOMS (GPU)}  & 10.40 (1×)      & \(4.68 \times 10^6\) (1×)     & N/A \\ \hline
\textbf{3D NAND (SLC)}    & 2.58 (4×)       & 949 (4.93E3×)  & 20.02 (1×) \\ \hline
\textbf{3D NAND (TLC)}    & 0.75 (13.9×)    & 763 (6.14E3×)  & 6.67 (3×) \\ \hline
\textbf{FeNOMS (PF\textsubscript{3}, $m$=1)} & 0.24 (43.9×)   & 187 (2.50E4×)  & 6.67* (3×) \\ \hline
\textbf{FeNOMS (PF\textsubscript{3}, $m$=4)} & 0.06 (175.7×)  & 46.9 (9.97E4×)  & 6.67* (3×) \\ \hline
\textbf{FeNOMS (PF\textsubscript{4}, $m$=4)} & 0.05 (224.1×)  & 37.1 (1.26E5×)  & 5.27\* (3.8×) \\ \hline
\end{tabular}
\vspace{-1mm}
\label{tab:comparison}
\begin{flushright}
\scriptsize *Only z-scaling for FeNOMS.
\end{flushright}
\vspace{-2mm}
\end{table}

Table~\ref{tab:comparison} highlights the performance metrics (latency, energy, and area) of various FeNOMS configurations compared to baseline HyperOMS (GPU)\cite{hyperoms} and 3D NAND (SLC and TLC) \cite{SLC3DNAND} ISP, all evaluated using HD-based workloads. SLC 3D NAND \cite{SLC3DNAND} achieves a 4× speedup over HyperOMS due to in-storage processing.  Transitioning to TLC reduces the required cells for storage capacity, enabling bitline scaling but increasing wordline delays. However, TLC requires seven sensing operations per wordline to read three bits per cell, introducing a bottleneck. FeNOMS addresses this challenge by leveraging the D-BAM method, which reduces the sensing operations to just two (UBC and LBC).

The conservative configuration of FeNOMS (PF\textsubscript{3}, $m=4$) achieves a 43× and 13× speedup over SLC and TLC 3D NAND, respectively, and delivers 176× higher performance than HyperOMS while maintaining minimal identification drop, as illustrated in Fig.~\ref{fig:heatmap}.
Additionally, FeNOMS offers 21× and 16× greater energy efficiency compared to SLC and TLC 3D NAND, respectively, while consuming five orders of magnitude less energy than HyperOMS.
FeNOMS also provides a 3× improvement in area efficiency over SLC 3D NAND, attributed to dimension packing. However, it shows no significant area advantage compared to TLC 3D NAND.
Nevertheless, the shorter FeNAND string length—reduced by a factor of \(k\)-enhances its scalability and makes it well-suited for high-density storage applications.

\subsection{FeNOMS: Design Space Exploration (DSE)}
\begin{figure}[h]
    \centering
    \includegraphics[width=0.48\textwidth,page=1]{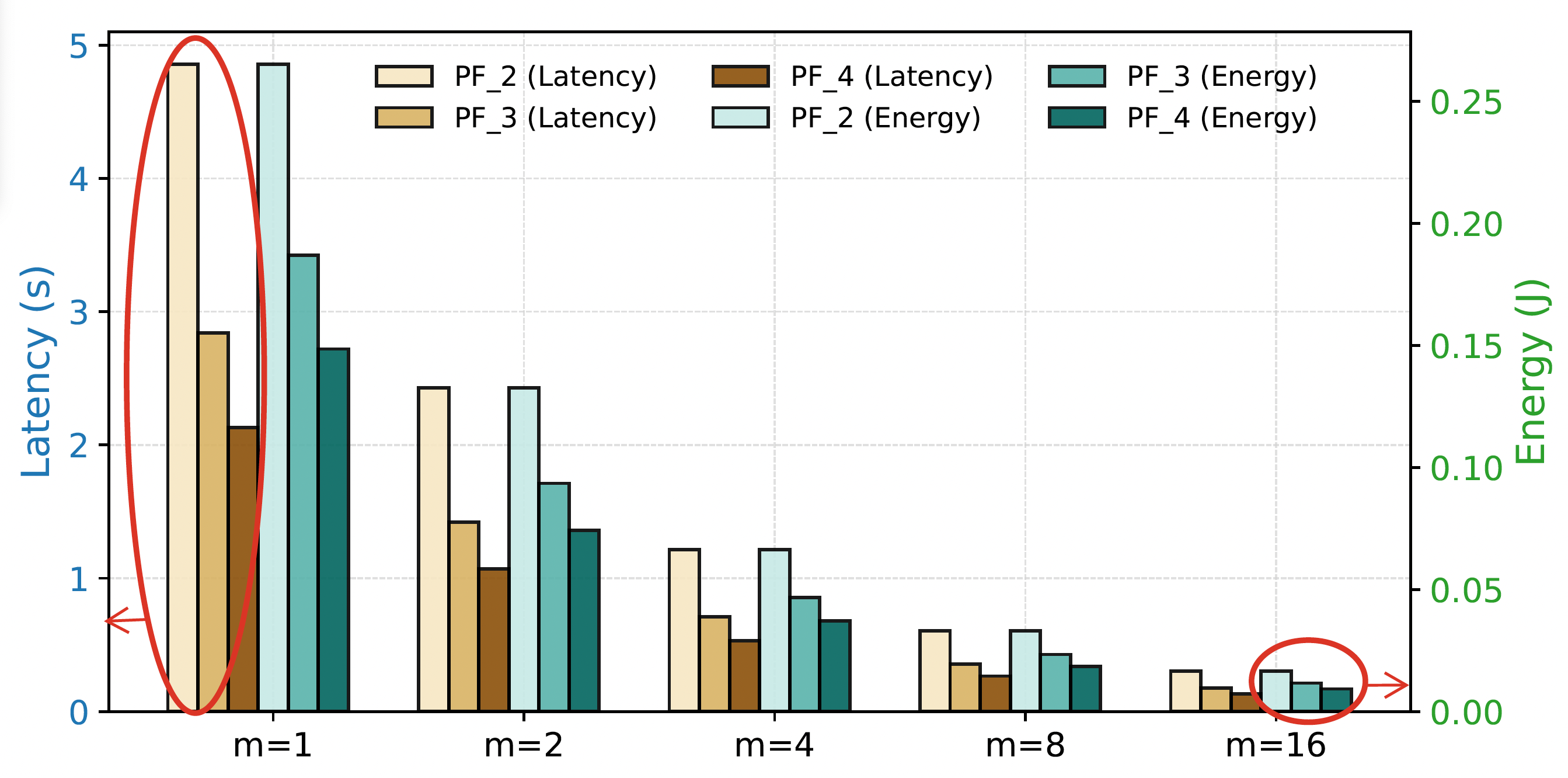}
    \caption{Design space exploration of FeNOMS configurations.}
    \label{fig:dse}
\end{figure}

Fig.~\ref{fig:dse} demonstrates the scalability and robustness of FeNOMS configurations with realistic high-density scenarios. We used prior work configuration of 512 wordlines to evaluate scaling TLC FeNAND\cite{512}.  We adjusted the peripheral circuit performance metrics to reflect the increased wordline count.  FeNOMS configurations are very efficient despite the non-linear area scaling associated with the peripheral circuit adjustments. Specifically, focusing on PF\textsubscript{3} with $m=4$, a 14$\times$ speedup in latency and an 11$\times$ improvement in energy efficiency are achieved compared to the baseline PF\textsubscript{2}, $m=1$, with less than a 10\% drop in identification rate. This showcases the adaptability of FeNOMS to improve  delay,  energy and area efficiencies dramatically by trading-off the accuracy gracefully.

\section{CONCLUSION}
This work presents FeNOMS, an in-storage processing architecture utilizing Ferroelectric NAND (FeNAND) for efficient OMS library searching. By integrating HDC with the novel Dual-Bound Approximate Matching (D-BAM) metric, FeNOMS overcomes limitations of conventional NAND architectures, achieving a 43× speedup and 21× higher energy efficiency compared to state-of-the-art 3D NAND methods. It maintains accuracy by tolerating non-idealities through a configurable margin, establishing FeNOMS as a scalable, energy-efficient solution for large-scale mass spectrometry. While we demonstrated the effectiveness of D-BAM for OMS search in this work, the same principle can be readily extended to a broad range of data search tasks across diverse data-intensive applications.

\section*{ACKNOWLEDGMENT}
This work was supported in part by PRISM and CoCoSys, centers in JUMP 2.0, an SRC program sponsored by DARPA (SRC grant number - 2023-JU-3135). This work was also supported by NSF grants \#2003279, \#1911095, \#2112167, \#2052809, \#2112665, \#2120019, \#2211386.

 \clearpage 
\bibliographystyle{IEEEtran}
\bibliography{FeNAND}

\end{document}